# Random noise can help to improve synchronisation of excimer laser pulses

Róbert Mingesz[1], Angéla Barna[2], Zoltán Gingl[1,*] and János Mellár[1]

[1]*Department of Technical Informatics, University of Szeged, Árpád tér 2, Szeged, 6720, Hungary*
[2]*Department of Experimental Physics, University of Szeged, Dóm tér 9, Szeged, 6720, Hungary*

## 1. Summary

Recently we have reported on a compact microcontroller-based unit developed to accurately synchronise excimer laser pulses (Robert Mingesz et al, Fluct. Noise Lett. 11, 1240007 (2012), DOI: 10.1142/S021947751240007X, http://arxiv.org/abs/1109.2632). We have shown that dithering based on the random jitter noise plus pseudorandom numbers can be used in the digital control system to radically reduce the long-term drift of the laser pulse from the trigger and to improve the accuracy of the synchronisation. In this update paper we present our new experimental results obtained by the use of the delay controller unit to tune the timing of a KrF excimer laser as an addition to our previous numerical simulation results. The hardware was interfaced to the laser using optical signal paths in order to reduce sensitivity to electromagnetic interference and the control algorithm tested by simulations were applied in the experiments. We have found that the system is able to reduce the delay uncertainty very close to the theoretical limit and performs well in real applications. The simple, compact and flexible system is universal enough to be used in various multidisciplinary applications as well.

## 2. Introduction

Precise synchronisation of events in time is essential in many experiments when time dependent behaviour is monitored. Very short time observations often aided by impulse lasers in exploring many physical, chemical, biological phenomena, pump-probe measurements, time-resolved spectroscopy [1-4]. A single trigger signal may generate multiple events including the laser shots and certain time delays appear at each individual component. Excimer laser operation itself may also need synchronising units to control X-ray pre-ionization and master oscillator-power amplifier setups [5]. Lasers exhibit a varying time delay therefore this must be controlled in order to maintain measurement accuracy. The hydrogen thyratron used in excimer lasers to activate the high voltage discharge has a certain switching time (so-called anode delay) which is subjected to a long-term drift caused by temperature changes. Since the laser operated at high frequency is heated up, the anode delay may be changed by a few tens of nanoseconds. A random jitter is also present in laser systems due to the gas discharge uncertainty. This shot-to-shot jitter is not predictable therefore the slowly changing part can be compensated only. The proper control of the delay needs precise time measurement and tuneable delay units [6,7]. Rather surprisingly the presence of random noise is not necessarily bad, sometimes adding noise can even improve performance. There are several applications today based on the constructive role of noise in various multidisciplinary fields. Examples include stochastic resonance that has special importance in biology [8-11] and dithering used in electrical engineering [12] and in image processing [13]. Recently we have developed a universal, simple microcontroller based delay controller unit that features flexible software control and precise time-synchronising hardware components [7]. Although the time resolution of the unit is limited to 10 ns, we could improve the timing accuracy well below this by utilizing the jitter noise of the laser and additional random dithering. As a result stability close to the theoretical limit could be achieved. Up to now the system's performance was tested by numerical simulations only, here we report on our experimental results of a laser pulse synchronisation application in a KrF excimer laser.

## 3. Experimental setup and delay controller operation

The principle of the control of the entire delay can be seen on figure 1. The rising edge of the external trigger impulse starts a programmable delay unit that activates the circuit of the thyratron. The high-voltage laser discharge is started by the thyratron signal and a laser pulse is generated. The desired delay is achieved by tuning the programmable delay; the total delay is the sum of this tuneable delay and the thyratron delay. The aim of the control is to minimize the error of the total delay, i.e. the difference between the total delay and the desired delay.

*Author for correspondence (gingl@inf.u-szeged.hu).

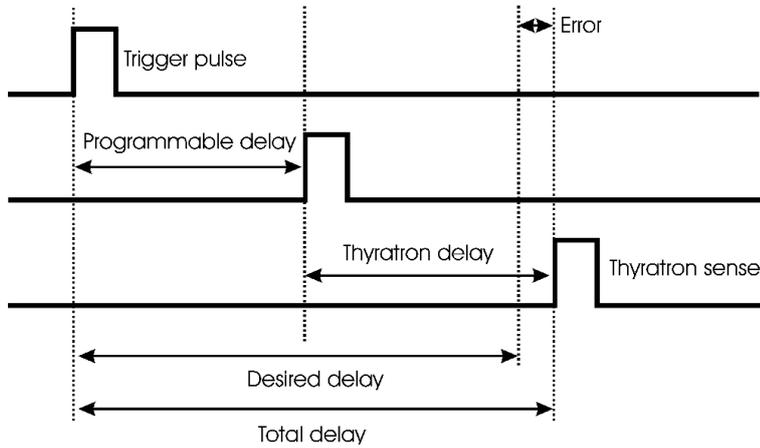

Figure 1. The principle of the control. The programmable delay can be used to compensate the changes in the thyratron delay and to achieve a total delay close to the desired value.

In the following we give a brief overview of the hardware components that were used to control the laser pulse timing. Note that only a smaller part of our more universal system was required therefore it is possible to build an optimized, even simpler unit in similar cases. The block diagram of the experimental setup and the simplified hardware of the delay controller unit are shown on figure 2.

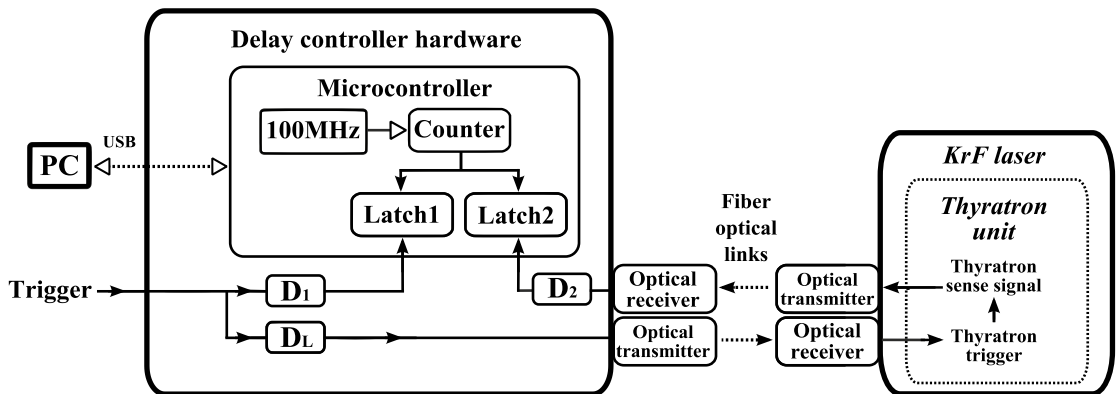

Figure 2. The block diagram of the experimental setup.

The delay controller unit is based on a C8051F120 microcontroller incorporating a 100MHz oscillator, a counter with two capture latches (Latch1 and Latch2) and a serial communication port. Three programmable delay units ($D_1, D_2, D_L$) are integrated on the controller unit and the KrF excimer laser is connected to the rest of the system via fibre optical links to reduce electromagnetic interference and noise. The whole system can be monitored by a host computer via a USB interface. The time diagram of the delay control operation is shown on Figure 3.

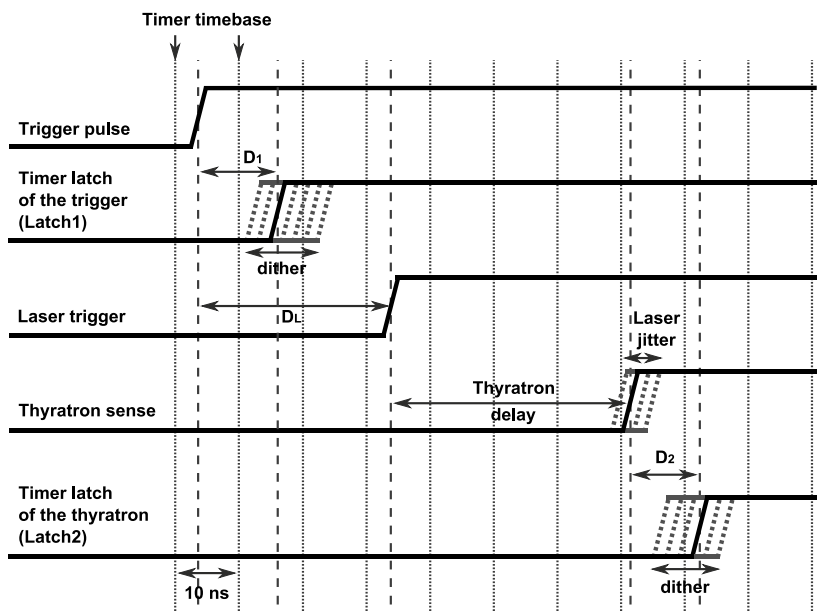

Figure 3. Time diagram of the delay control operation.

The programmable delay unit $D_L$ can delay the triggering of the thyratron that activates the laser; this is used to achieve the desired overall delay. The delay time measurements are performed by time-to-digital converters implemented by the microcontroller's integrated programmable counter array (PCA). Its 16 bit main counter is driven by the 100MHz oscillator in order to achieve a resolution of 10 ns. The two independent latches can capture the actual value of this counter upon the rising edge of the connected signals. The external trigger time instant delayed by the programmable delay unit $D_1$ is measured by Latch1, while the thyratron sense signal delayed by $D_2$ is captured by Latch2. The units $D_1$ and $D_2$ are required to optimize the quantization of the total delay time of the system by adding random dither to improve the resolution of the time measurement as it was described in the externally triggered operation chapter of our initial article [7]. Subtracting the values stored in Lacth2 and Latch1 gives the digitized total delay. Since due to the dithering and laser jitter this value is fluctuating, adaptive averaging can be used to improve accuracy significantly, see the control algorithm section in [7]. The averaging time must be small enough in order to minimize the effect of the long-term drift of the thyratron delay. If the total delay is measured, it can be set to the desired value by tuning $D_L$.

## 4. Experimental results

In order to reduce electromagnetic interference and to avoid possible ground loops the controller hardware and the KrF excimer laser trigger and thyratron sense signal were connected using fibre optic cables, optical receivers (HFBR 2521) and transmitters (HFBR 1521). These elements introduce excessive jitter therefore first we have evaluated the performance without the use of the KrF excimer laser. The desired delay was set to 1000 ns, and the jitter was measured using the time-to-digital conversion method described in the previous section. We have found that the standard deviation of the jitter of the timing circuitry was 2.1 ns. In order to measure the full jitter of the system the delay control was switched off. The full delay between the laser trigger and the thyratron sense signals was measured to be 1.36 µs with a jitter of 10 ns. Due to the statistical independence the full jitter is the root mean square of the laser jitter plus the jitter of the timing circuitry. In our case this means that the contribution of the timing circuitry was below 2.2 % of the full jitter. Lower jitter lasers may need even better performing optical receivers and transmitters. As a rule of thumb, the contribution to the total jitter can be kept below 5 % if the jitter of the system is less than the one third of the laser jitter. After the testing the control circuit was tested with the KrF laser system. During the experiments a pulse shape time dependent desired delay was set as depicted in figure 4.

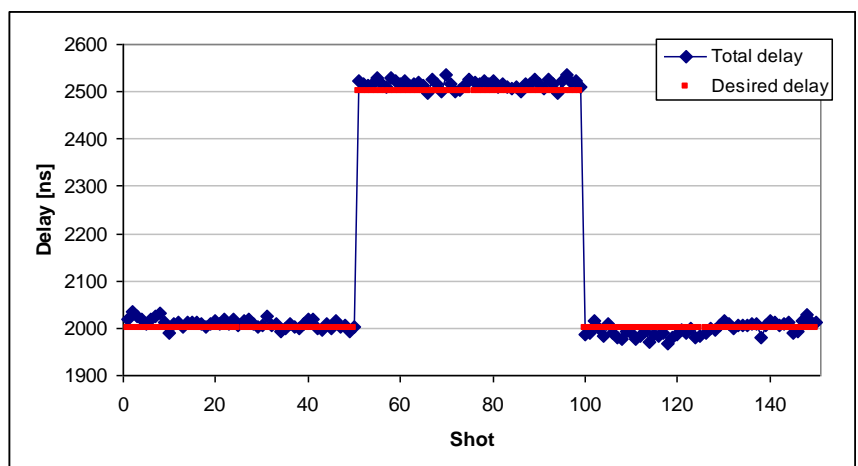

Figure 4. The regulated total delay and the desired delay as function of the laser shots.

After 50 shots the initial value of 2 µs was changed to 2.5 µs and after next 50 shots it was set again to 2 µs. Table 1 summarizes the measurement results for the three parts, in each case the number of averages was 50.

| Desired delay | Measured delay | Jitter |
| --- | --- | --- |
| 2,0 µs | 2.010±0.003 µs | 10 ns |
| 2,5 µs | 2.517±0.003 µs | 9 ns |
| 2,0 µs | 1.997±0.004 µs | 13 ns |

Table 1. Desired and measured delay and jitter for the three different parts of the laser delay regulation.

The measurement results show clearly that high accuracy can be achieved with the developed control system while the additional jitter due to the control system is negligible.

## 5. Conclusion

In addition to the results reported in our initial article in this update paper we have demonstrated the performance of our compact microcontroller based delay control system driving a KrF excimer laser. We have added fibre optic links to the original system to reduce its sensitivity to interference and noise and we have shown that the time synchronisation accuracy of the system can be maintained. Our experimental results confirm that the long-term drift of the laser delay can be removed by utilizing the shot-to-shot jitter noise and added random dither with a simple and compact system. We have shown that the regulation can keep the accuracy close to the theoretical limit determined by the jitter of the laser. Lasers controlled by such system and algorithm can be used in many multidisciplinary applications where timing accuracy is important. Note also that the very compact and universal system can play constructive role in general time-to-digital conversion and precision time synchronisation solutions in various fields of science and engineering.

**Acknowledgments**
The technical support of the KrF laser system in the HILL Laboratory led by S Szatmari is gratefully acknowledged. We thank J. Bohus, S. Szanto and G Szabo for their valuable technical assistance.

**Competing Interests**
We have no competing interests.

**Authors' Contributions**
RM and ZG conceived of the study, RM developed and implemented the control algorithm and performed the simulations, ZG worked out the principle of the measurement and control circuit and identified the components, JM drawn the schematic, designed the circuit board, built and tested the circuit, AB and ZG designed and coordinated the experimental study, AB integrated the control circuit with the laser, carried out the experiments, made the measurements, analyzed the measurement data and carried out the lab work, AB and ZG drafted and wrote the manuscript. All authors gave final approval for publication.